\documentclass[aip,apl,reprint]{revtex4-1}
\usepackage{graphicx}
\usepackage{amssymb,amsmath,verbatim}
\usepackage{natbib}
\usepackage{epstopdf}
\usepackage{array}

\newcommand{\oneover}[1]{\frac{1}{#1}}

\newcommand{\grad}{\nabla}
\newcommand{\lapl}{\Delta}

\renewcommand{\div}{\nabla \cdot}

\begin{document}
\title{Development of Knife-Edge Ridges on Ion-Bombarded Surfaces }
\author{Miranda Holmes-Cerfon}
\affiliation{Harvard School of Engineering and Applied Sciences}
\affiliation{Kavli Institute for Bionano Science and Technology,
29 Oxford St.  Cambridge, MA 02138}
\author{Wei Zhou}
\affiliation{Harvard School of Engineering and Applied Sciences}
\affiliation{Permanent address: School of Mechanical and Aerospace Engineering, Nanyang Technological University, Singapore 639798, Singapore}
\author{Andrea L. Bertozzi}
\affiliation{Department of Mathematics, University of California, Los Angeles, Los Angeles, CA 90095-1555}
\author{Michael P. Brenner}
\affiliation{Harvard School of Engineering and Applied Sciences}
\affiliation{Kavli Institute for Bionano Science and Technology,
29 Oxford St.  Cambridge, MA 02138}
\author{Michael J. Aziz}
\affiliation{Harvard School of Engineering and Applied Sciences}

\begin{abstract}
We demonstrate in both laboratory and numerical experiments that ion bombardment of a modestly sloped surface can create knife-edge like ridges with extremely high slopes. Small pre-fabricated pits expand under ion bombardment, and the collision of two such pits creates knife-edge ridges.    Both laboratory and numerical experiments show that the pit propagation speed and the precise shape of the knife edge ridges are universal, independent of initial conditions, as has been predicted theoretically.
These observations suggest a novel method of fabrication in which a surface is pre-patterned so that it dynamically evolves to a desired target pattern made of knife-edge ridges.
\end{abstract}

\maketitle

The efficient fabrication of ever-smaller structures is one of the major challenges of 21st century science and engineering. Ion bombardment has emerged as a promising candidate to create patterns on surfaces \cite{sigmund69,sigmund73,bradley88,chason}. One method uses a focused ion beam to micro-machine sharp features directly\cite{vasile97,adams03,li01,stein02} -- this allows for detailed control of the shape of the features but, as a serial writing process, is too time-consuming to pattern large areas.
Another method is to bombard a surface uniformly, which can excite linear instabilities that grow into patterns such as quantum dots \cite{facsko99,frost00,cuenat05}. This is less costly, but can only create structures as small as the smallest linearly unstable wavelength, with steepnesses limited due to saturation of the linear modes. To overcome these limitations one would like to create steep, sharp structures spontaneously, by exploiting the dynamical processes that underlie surface evolution under ion bombardment\cite{white}.

\begin{figure}[h!]
\begin{center}
\includegraphics[width=0.85\linewidth]{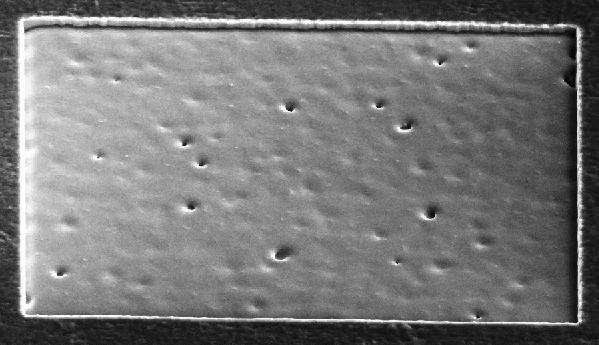}\\
\includegraphics[width=0.85\linewidth]{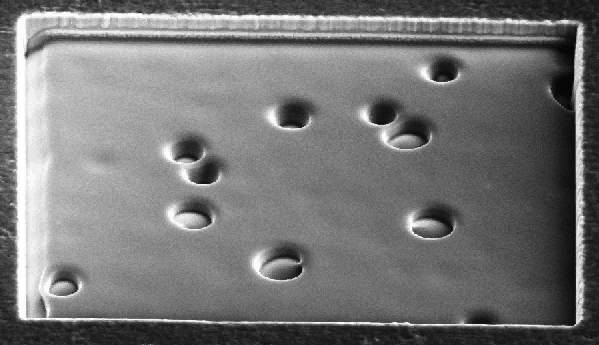}\\
\includegraphics[width=0.85\linewidth]{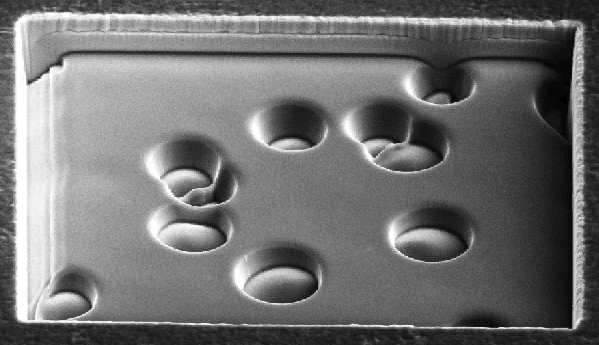}\\ 
\end{center}
\caption{Surface evolution of a magnesium alloy under uniform irradiation by a focused ion beam, after 5, 12, 19 minutes. The surface initially contained small holes that grew from scratches. The imaged region is 30$\mu$m, viewed at 52$^\circ$.
}\label{fig:expt}
\end{figure}

Recently we proposed a scenario for creating very sharp features on ion bombarded surfaces, by starting with a surface that is pre-patterned to have modest
slopes on the macroscale \cite{MHC2012}.  Under ion bombardment, our theoretical calculations demonstrated that if the initial slope is in the right range, the structures would spontaneously evolve to knife-edge-like ridges, with extremely high slopes, and high radius of curvature.  Both the final slope and radius of curvature are independent of initial conditions, and depend only on the shape of the curve describing sputter yield (atoms out per incident ion) vs. incidence angle.  Here we demonstrate the formation of knife edge ridges in experiments.
Our experiments show that uniformly irradiated small pits expand outward, developing steep sides with uniform slopes. When two pits collide, the front evolves to a sharp, knife-edge-like structure with features on a scale much smaller than any contained in the initial conditions. Numerical simulations of the classical macroscopic equations show remarkably similar dynamics.
Both experiment and simulations show that the pit propagation speed and the precise shape of the knife edge ridges are universal, independent of initial conditions, as predicted theoretically.
These dynamics can be understood by a theoretical analysis of the equations in which the knife-edge structure arises as a particular traveling wave solution with a large basin of attraction. Because there is only one such solution, the dynamics are relatively insensitive to the initial conditions and a pre-patterned surface will evolve to uniform knife-edge ridges with the same slopes and radii of curvature.
If one can learn to control the location of the ridges by solving an inverse problem, one could potentially make a desired target pattern out of the knife-edge ridges.

\begin{figure}
\begin{minipage}{0.48\linewidth}
\includegraphics[width=1\linewidth]{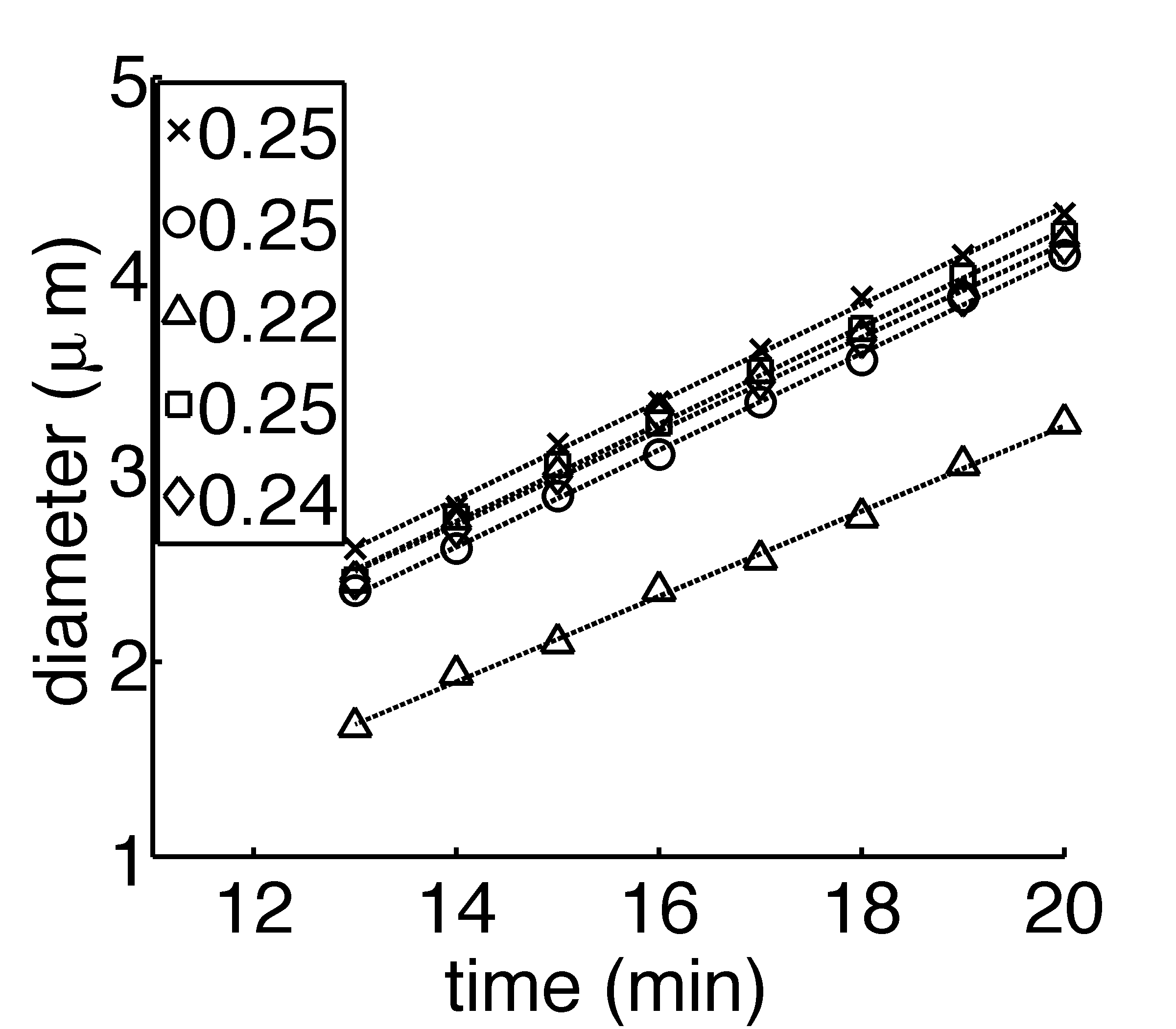}
\end{minipage}
\begin{minipage}{0.48\linewidth}
\includegraphics[width=1\linewidth]{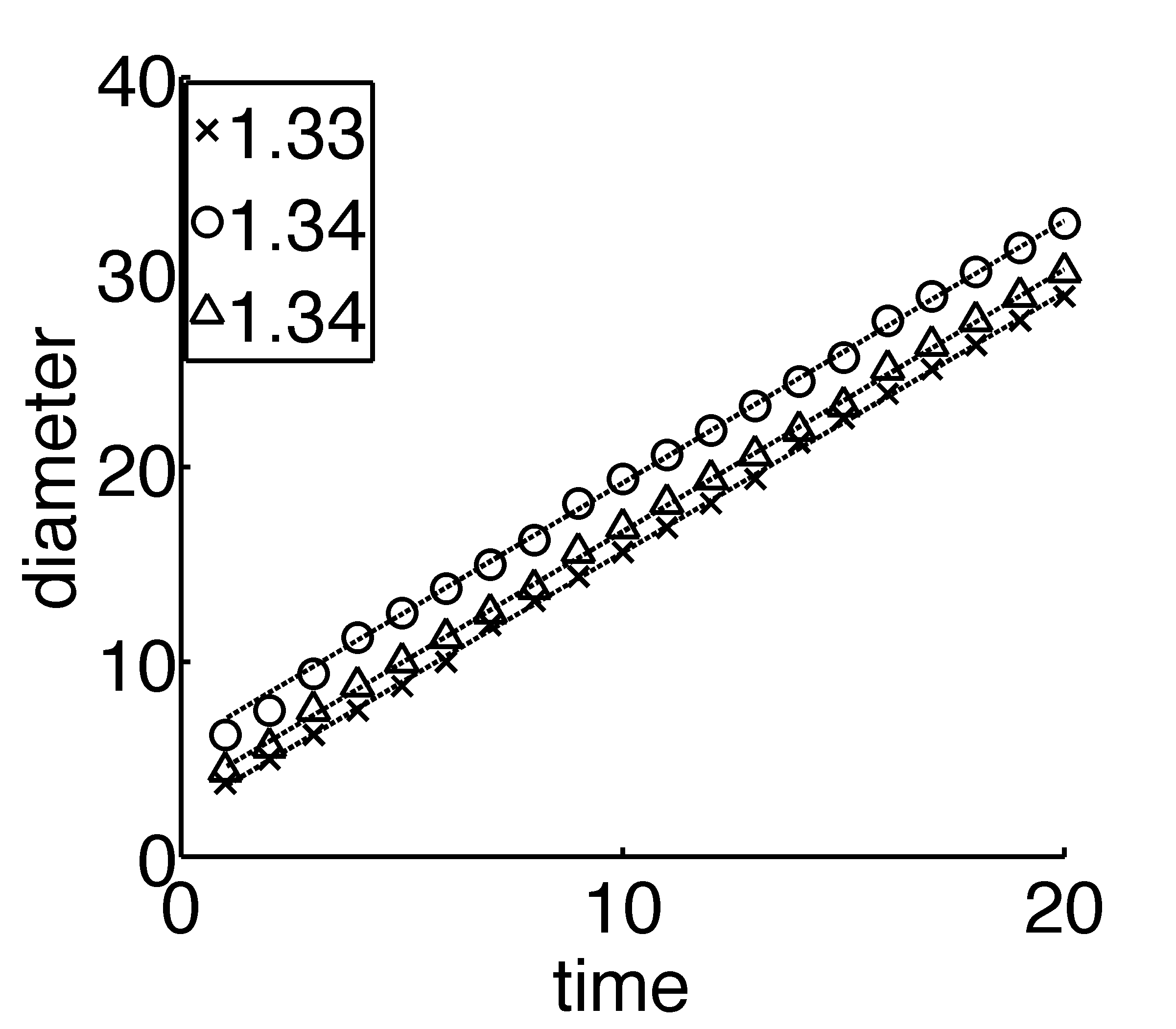}
\end{minipage}
\caption{Diameter of pits in experiment (left) and simulations (right), as a function of time. Each marker represents a different pit (expts) or initial condition (sims); experiments are accurate to $\pm 0.15 \mu$m. Best-fit lines are dashed and legend indicates their slopes.
The initial conditions for the simulations were:
(cross) $h(x,y) = -6e^{-(x^2+y^2)}$, (circle) $h(x,y) = -10e^{-(x^2+y^2)/4}$, (triangle) $h(x,y) =  -6e^{-(x^2+y^2)/r(\theta)^2}$ with $r(\theta) = 1+0.5\sin(4\theta)$, $\tan \theta = x/y$. 
}\label{fig:slopes}
\end{figure}


\begin{figure}
\includegraphics[width=0.6\linewidth]{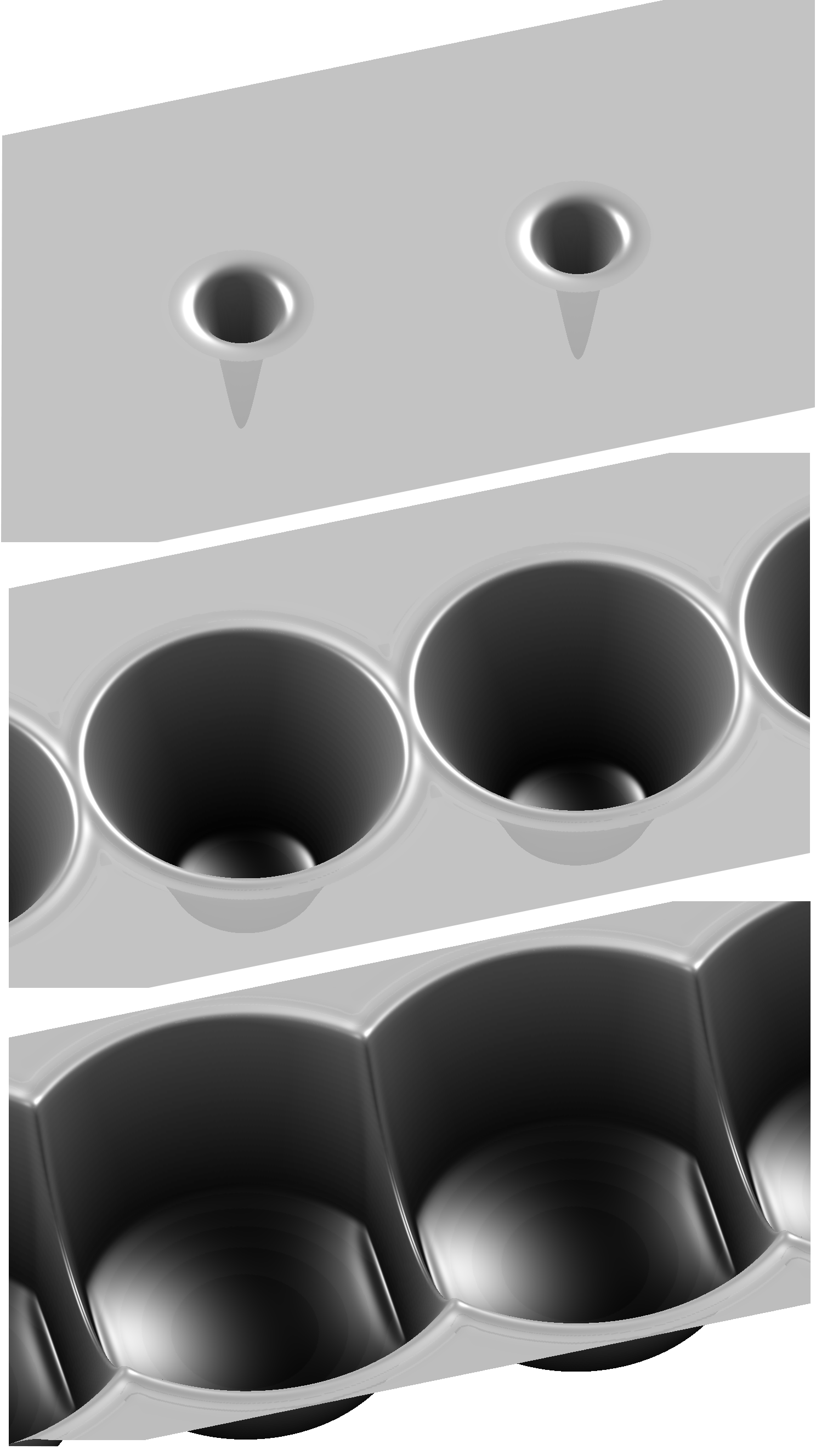}
\caption{Numerical simulations of Eq. \eqref{eq:h} with yield function Eq. \eqref{eq:Y}, $B_0 = 1/100$, 
at times $t=$ 0, 11, 19. The initial condition was $h(x,y) = h_0e^{\frac{-(x^2+y^2)}{2\sigma^2}}$.}\label{fig:simpits}
\end{figure}

\begin{figure}
\begin{center}
\includegraphics[width=0.48\linewidth]{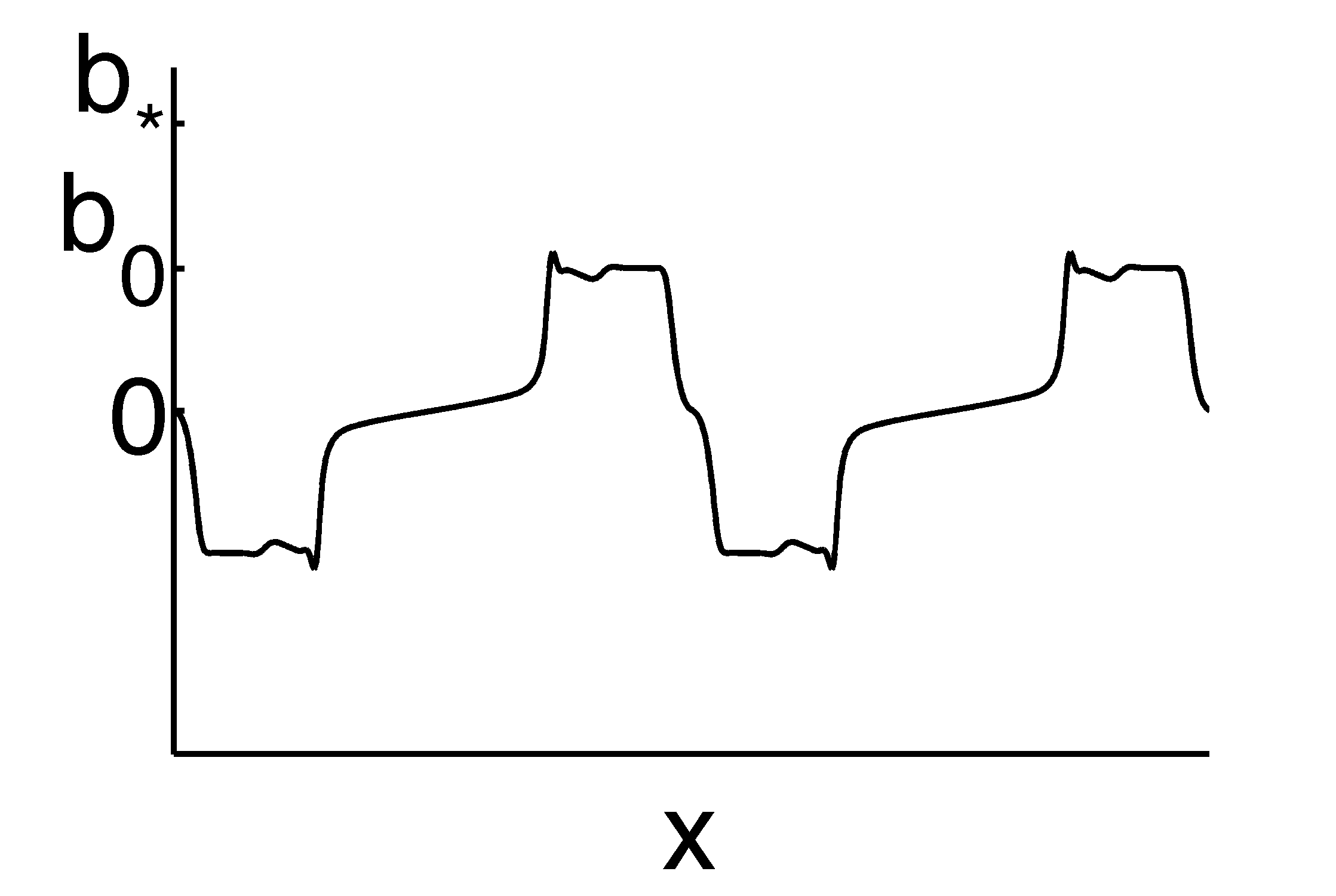}
\includegraphics[width=0.48\linewidth]{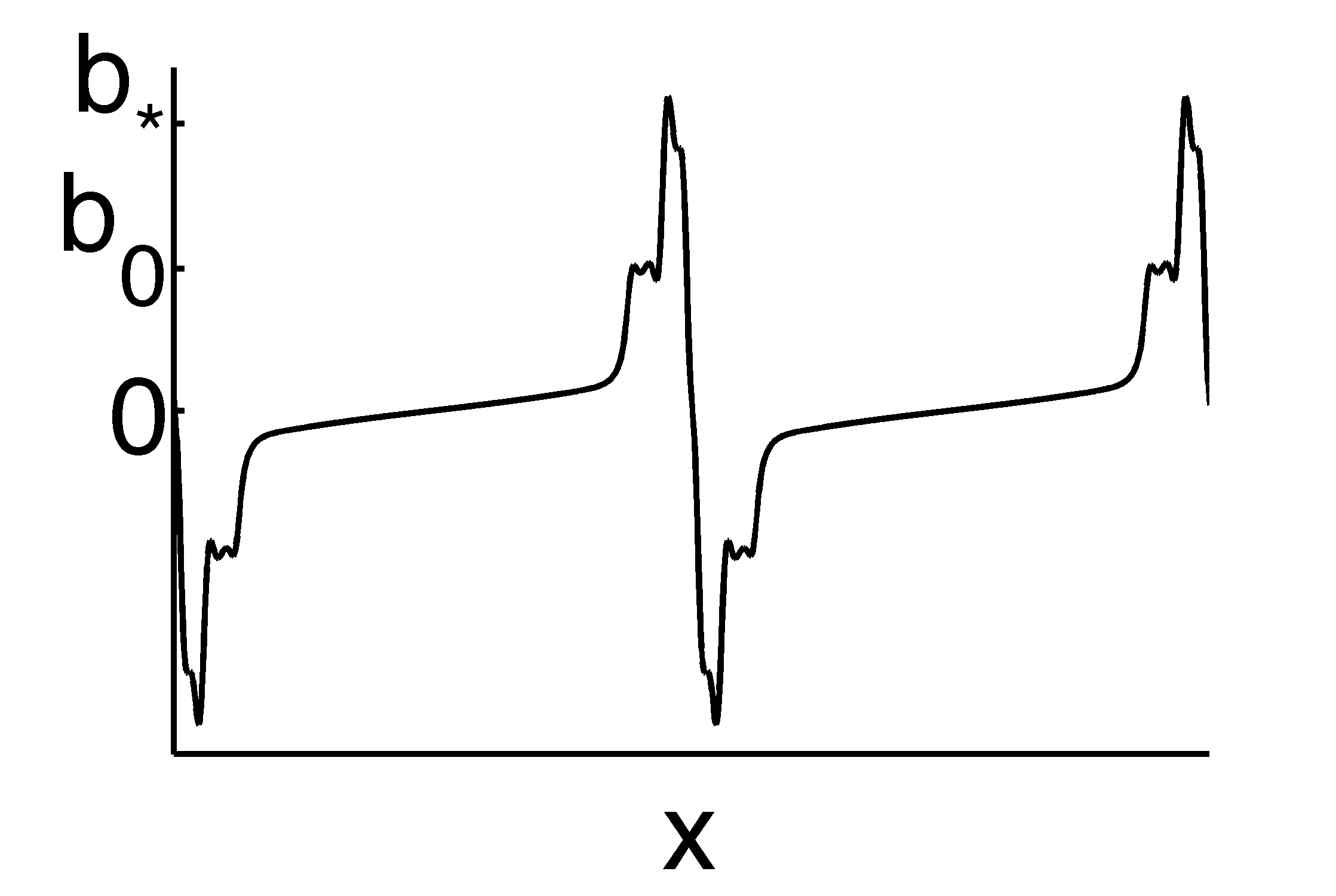}\\
\includegraphics[width=0.7\linewidth]{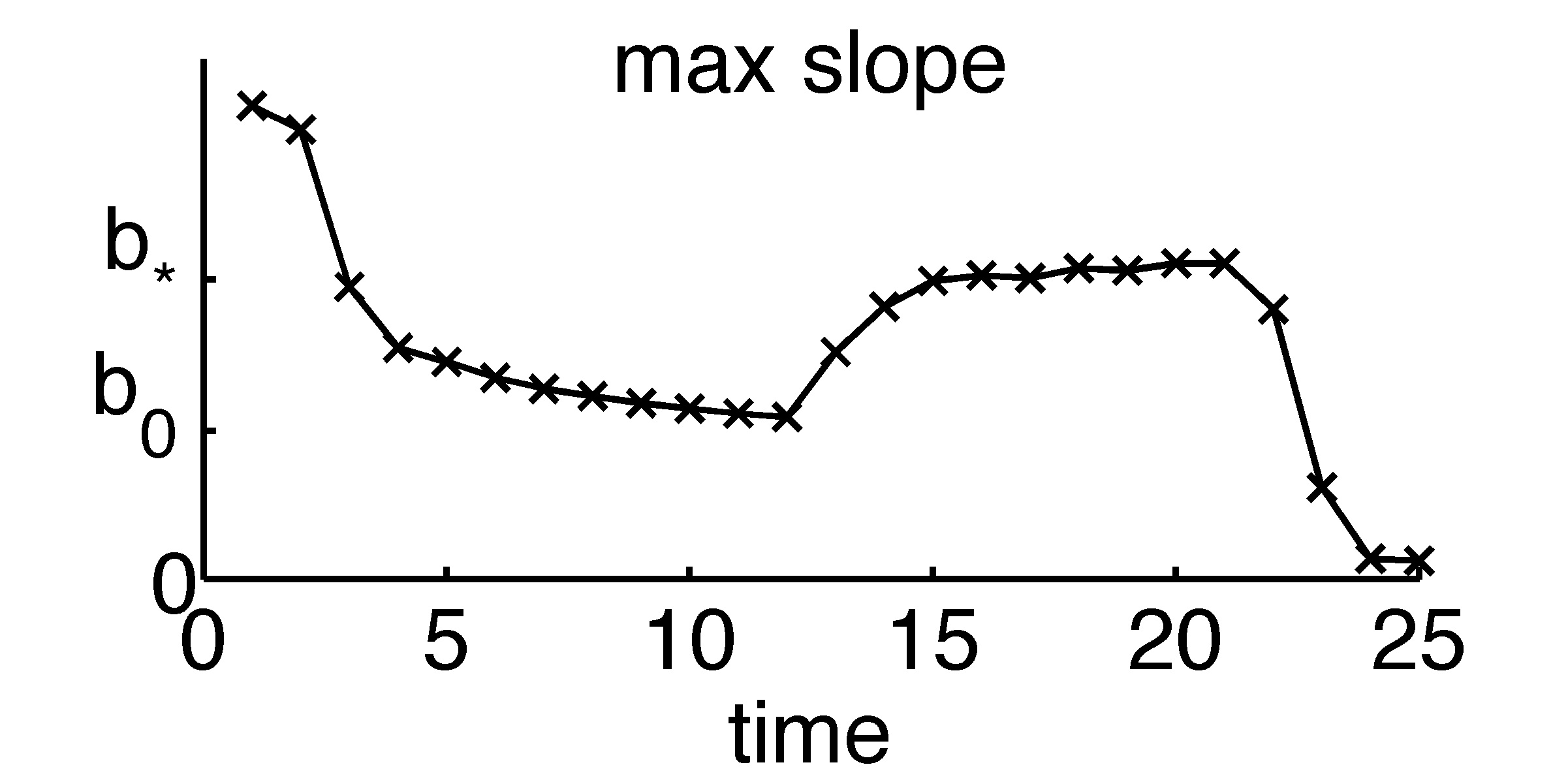}
\end{center}
\caption{Top: slope $h_x$ through a horizontal cross-section $y=0$, at the center of the simulations in Figure \ref{fig:simpits}, at time $t=11$ (left), $t=15.5$ (right). The ordinate axis indicates slopes $b_0=2.3$, $b_*=4.7 $ that act as dynamical attractors for the one-dimensional traveling wave equation.  Bottom: maximum slope $|h_x|$ as a function of time. An initial transient period during which the narrow initial condition adjusts to the undercompressive shock $b_0$, is followed by collision of shocks where the maximum slope jumps to $b_*$.
}\label{fig:simpitsslope}
\end{figure}

Our experiments were performed on an ingot of a magnesium AZ91D alloy, with nominal alloying element content 9 wt.\% Al and 1 wt.\% Zn; however, the experiments were confined to the alpha (aluminum-poor) phase of a two-phase mixture. The surfaces were uniformly irradiated using a FEI dual beam Focused Ion Beam (FIB) - Scanning Electron Microscope (SEM) delivering 30 keV Ga$^{+}$ to the surface in a background pressure of 1.4$\times 10^{-6}$ mbar at room temperature. The incident ion beam was parallel to the surface normal and the ion beam current was 3 nA.  The beam was rastered in a boustrophedonical scan across a pre-defined region of the sample surface, during which time the beam would dwell at each discrete location for 0.1 $\mu$s and then move rapidly to an adjacent location. Separation between adjacent locations was set to nominally 50\% overlap, which in this case meant a 75 nm center-to-center spacing for a 150 nm diameter beam.  The current profile within the beam is believed to be roughly Gaussian.
The irradiation was interrupted periodically so that the irradiated surfaces could be observed using in-situ SEM from both normal incidence and tilted 52 degrees.

The surface topography was initially irregular due to metallographic polishing scratches, and some of these irregularities initiated small holes in the surface (Figure \ref{fig:expt}, top.) Most of the holes then decayed to a flat surface, but certain holes developed into pits that continuously expanded. What is notable is the pits appear to be identical: they expand at the same rate, and have sides of the same slope. We measured the diameters of five different pits as a function of time from a normal view of the surface. These changed at an average rate of $0.24 \mu$m/min, with all rates lying within $0.02 \mu$m/min of the average,  well within the resolution of our measurement of the diameter  (Figure \ref{fig:slopes}, left).
We are not able to quantitatively compare the slopes, but qualitative examination of Figure \ref{fig:expt} middle, bottom suggests they are also very similar.

When two pits collided they created very steep, sharp ridge-like structures.
There are three examples of such ridges in Figure \ref{fig:expt} (bottom). These are notable because the length scale that is created by the collision is much smaller than any contained in the initial condition. Qualitatively, the steep ridges have similar slopes in each of the three cases.

We now turn to numerical simulation of these structures,
using a partial differential equation governing the evolution of the surface height $h(x,y,t)$ on the macroscale:
\begin{equation}\label{eq:h}
h_t + R(b)  + B_0\div\left(\oneover{\sqrt{1+b^2}} \grad \kappa\right) = 0.
\end{equation}
This equation is derived from the widely-used Sigmund theory of sputter erosion \citep{sigmund73}, by expanding the sputter integral for surfaces whose curvature is much smaller than the lateral scale over which an ion deposite its kinetic energy \cite{chen2005, aziz2006}.
Here $R(b)=R_0\sqrt{1+b^2}Y(b)$ is the the average velocity of erosion of the surface as a function of its slope $b = |\grad h|$ (or equivalently the angle of the incoming ion beam), obtained from the yield function 
$Y(b)$  by multiplying by a dimensional factor $R_0$ and a geometrical factor. The fourth-order term with magnitude $B_0$ is a function of the surface curvature $\kappa = \div\left(\oneover{\sqrt{1+b^2}} \grad h\right)$, and models additional smoothing effects such as Mullins-Herring surface diffusion \citep{mullins1959,herring1950} or ion-enhanced viscous flow confined to a thin surface layer \citep{umbach2001}.
We neglect the second-order (curvature) terms that are often included \citep{bradley88, davidovitch2009}, as the dynamics we are interested in occur when these are small.

We have developed an efficient, stable
method to solve Eq. \eqref{eq:h} in two dimensions. Solving such higher-order nonlinear equations in multiple dimensions is a generally a challenge -- explicit methods impose severe restrictions on the time step $\Delta t$ for the scheme to be stable ($\Delta t < O(\Delta x)^4$, where $\Delta x$ is the grid spacing\cite{strikwerda,bertozzi93}), while fully implicit schemes, that are unconditionally stable, require solving a difficult nonlinear problem at each time step. We overcome these difficulties by adding a fourth-order linear term to the equation that we treat implicitly, while treating the nonlinear parts of the equation explicitly \cite{vollmary2003}. Specifically, our scheme takes the form
\begin{multline}\label{eq:scheme}
h^{j+1} + M\Delta t \lapl^2 h^{j+1} =\\
h^j  -\Delta t\left( R(|\grad h^j|) - N(|\grad h^j|) + M\lapl^2 h^j  \right)
\end{multline}
where $h^j(x,y)$ is the solution at time $j\Delta t$,  $N(b)$ is the fourth-order nonlinear term, and $M>0$ is a constant that we are free to choose. Analysis for similar equations \cite{gao2011, carola2011, bertozzi2011} has shown the scheme is stable irrespective of the grid spacing provided $M$ is large enough.
However, with fixed timestep the accuracy decreases if $M$ is too large. We found a good balance between stability and accuracy with $M=1$.

To apply \eqref{eq:scheme}, the right-hand side is evaluated explicitly using centered finite differences for the spatial derivatives, the result is converted to Fourier space using periodic boundary conditions, and  $h_{j+1}$ is found in Fourier space by solving the linear inversion problem.

We perform simulations using a yield function of the Yamamura form  \cite{yamamura1983}
\begin{equation}\label{eq:Y}
\frac{Y(\theta)}{Y(0)} = (\cos\theta_{opt})^{-f} \exp\{ -\Sigma((\cos\theta_{opt})^{-1} - 1)\},
\end{equation}
where the parameters are $f=2.36$, $\theta_{opt} = 69.5$, and $\Sigma=f\cos\theta_{opt}$. Yamamura has shown that a great many experimentally measured yield functions can be represented in this form, by fitting for $f$, $\theta_{opt}$. Our theory (to be described) shows that the qualitative features of the dynamics are robust to changes in these parameters so we chose these for numerical convenience; we do not attempt quantitative comparison as the yield function for the experiments is unknown.

We ran two kinds of simulations.
First we looked at the formation of pits, by initiating the surface with individual small holes. For small perturbations the pits expand slowly and decay without creating smaller length scales. However, for large enough perturbations, the pits evolve to a circular crater whose sides have a uniform, steep slope, that expands outwards with a constant velocity. The slopes of the sides, and the speed of propagation, are fixed numbers, independent of the form of the initial perturbation.  Figure \ref{fig:slopes} (right) illustrates the constant speed for three different initial conditions.

Next, to investigate how pits collide, we initiated the surface with two nearby pits. These expand, and when they collide, they form a ridge whose sides are very steep -- much steeper than the sides of the crater. The slope of the sides is always the same, regardless of the  initial conditions. Figure \ref{fig:simpits} shows the surface evolution for one choice of initial condition, and
Figure \ref{fig:simpitsslope} (bottom) shows the maximum slope as a function of time for this simulation. The plateau from time 13--20 corresponds to the slope of the ridge, and is the same height for a broad class of initial conditions.

The simulations are striking because of their remarkable resemblance to the experiments.
Two notable features occur in both:
(1) Craters expand with a constant slope and velocity;
(2) When craters collide, they create a very sharp knife-edge-like ridge, with steeper slopes than those originally on the surface. In both cases the slopes are universal, independent of the initial condition.

These features were predicted by a  recently-developed theory\cite{chen2005,MHC2012}, as the consequence of the unusual type of traveling wave solutions that occur in the governing equations (1).
 When a pit is large enough, the crater rim is locally nearly straight, and can be well approximated by a traveling wave that is invariant in one horizontal direction. Therefore we look for traveling wave solutions to the one-dimensional equation. As shown in Holmes-Cerfon et. al \cite{MHC2012}, the slope $b=h_x$ can propagate as a traveling wave provided the slopes in the far-field are held constant, so we look for solutions $b = S(x-ct)$ to
 \begin{equation}\label{eq:S}
c(S-b_r) - (R(S) - R(b_r)) = B_0\left( \oneover{\sqrt{1+S^2}}\left(\frac{S'}{(1+S^2)^{3/2}}\right)'\right)' ,
\end{equation}
with boundary conditions $S(+\infty) = b_r$, $S(-\infty) = -b_l$, $S'(\pm \infty) = 0$.
This equation is obtained from \eqref{eq:h} by differentiating once, and then integrating from $+\infty$ to $x-ct$, while the speed $c$ is determined by conservation of mass to be
$ c =  \frac{R(b_r) - R(b_l)}{b_r-b_l}$.
Only certain pairs $(b_l, b_r)$ yield a solution, and this fact is crucial for determining the dynamics.

Ahead of the crater rim the surface is flat, so we look for solutions with boundary condition $b_r=0$. 
The theory shows that there are solutions $(b_l,0)$ for all $b_l$ less than a critical slope $b_0^c$ depending on the yield function. Above this critical slope there is exactly one boundary value yielding a solution: $b_l = b_0$, where $b_0$ is a number that again depends on the yield function.
This solution is isolated and serves as an attractor for the dynamics in the following sense:
if a one-dimensional surface is patterned initially to contain slopes greater than $b_0^c$, then these slopes will evolve spontaneously to the traveling wave connecting $b_0$ to $0$ as the sloped region propagates into the flat far-field\cite{MHC2012}.

This explains the first observation. When the initial perturbation is large enough, it evolves to a crater whose rim propagates outwards with speed $c$ corresponding to the discrete traveling wave $(b_0,0)$, and whose sides therefore have slope $b_0$. Figure \ref{fig:simpitsslope} (left) shows the slope $b$ through a horizontal cross-section  of the simulated craters just before they collide, which clearly shows the slope of the sides is a constant, uniform value $b_0$.

What happens when pits collide? The simulations suggest a symmetry so we look for solutions with boundary conditions $(b_l, -b_l)$. Again there is a threshhold determining the behaviour: when $b_l < b_*^c$ for some yield-function-dependent number $b_*^c$, there is always a solution, but when $b_l>b_*^c$, there is exactly one solution: $b_l = b_*$. This solution corresponds to a ridge with very steep sides and a small radius of curvature at the tip.
Simulations\cite{MHC2012} showed that this solution will evolve from two nearby regions with slopes of opposite signs, provided both have (not necessarily equal) magnitudes $>b_*^c$. Since $b_0 > b_*^c$, we predict that colliding pits will evolve to the knife-edge ridge.

Indeed, our numerical simulations confirm this -- Figure \ref{fig:simpitsslope} (right) plots the slope $b$ through a cross-section in the center of the colliding pits. The step-like appearance captures both of these traveling waves: the first step at $b_0$ is the slope of the original crater sides, and the second step at $b_*$ is the knife-edge ridge.
At later times (not shown) there is only one step, at $b_*$, as the crater sides have entirely evolved to the knife-edge.

For the yield function in our simulations the numerical values  are $b_0^c = 1.26$, $b_0 = 2.3$, $b_*^c = 1.28$, $b_* = 4.7$ -- but the values of the dynamical attractors increase with $\theta_{opt}$, and for certain materials we predict ridges with slopes of $b_*=30$ or more \cite{MHC2012}. Therefore we can create very sharp features by choosing an energy level or material that gives the desired values.

We have shown that the sharp, small-scale structures observed in our experiments and numerical simulations can be explained through the set of traveling wave solutions to the governing macroscopic equations. It is notable that only the macroscopic mechanisms of erosion and smoothing are required to instigate the observed features. Of course, additional small-scale physics 
may help to explain some of the qualitative \emph{differences} between the experiments and simulations, such as the curvatures of the pit bottoms. 
Indeed, we hypothesize that the experimental geometry may be significantly influenced here by multiple scattering effects, where ions incident on the pit wall and forward-scattered, as well as forward-sputtered atoms from the pit wall, may contribute to enhanced erosion along the perimeter of the pit bottom.

The theory predicts that two traveling wave solutions, both with steep slopes, control the dynamics over a wide range of initial conditions. What is potentially useful about these solutions is that they arise spontaneously from smaller slopes -- therefore we don't need to start with steep, small-scale structures in order to create them; these are created by the dynamics. This suggests a potential self-organizing principle for fabricating small-scale features on a surface,  by pre-patterning the surface on the macroscale so that it evolves to a structure built of small-scale ridges.
One is then interested in the inverse problem: to find an easily-achievable initial patterning of the surface, so that it evolves under uniform irradiation to a target small-scale pattern.

\acknowledgements{
The research of W.Z. was supported by grant RG79/98 provided by Nanyang Technological University. The research of M.J.A. was supported by DOE grant DE-FG-02-06ER46335.
MPB, MHC were funded by the National Science Foundation
through the Harvard Materials Research Science and Engineering Center
(DMR-0820484),  the Division of Mathematical Sciences (DMS-0907985)
and the Kavli Institute for Bionano Science and Techology at Harvard University.
AB, MHC were supported by NSF grant DMS-1048840 and UC Lab Fees Research Grant 09-LR-04-116741-BERA}


\bibliography{ShockBib.bib,NumericsBib.bib}

\begin{thebibliography}{26}%
\makeatletter
\providecommand \@ifxundefined [1]{%
 \@ifx{#1\undefined}
}%
\providecommand \@ifnum [1]{%
 \ifnum #1\expandafter \@firstoftwo
 \else \expandafter \@secondoftwo
 \fi
}%
\providecommand \@ifx [1]{%
 \ifx #1\expandafter \@firstoftwo
 \else \expandafter \@secondoftwo
 \fi
}%
\providecommand \natexlab [1]{#1}%
\providecommand \enquote  [1]{``#1''}%
\providecommand \bibnamefont  [1]{#1}%
\providecommand \bibfnamefont [1]{#1}%
\providecommand \citenamefont [1]{#1}%
\providecommand \href@noop [0]{\@secondoftwo}%
\providecommand \href [0]{\begingroup \@sanitize@url \@href}%
\providecommand \@href[1]{\@@startlink{#1}\@@href}%
\providecommand \@@href[1]{\endgroup#1\@@endlink}%
\providecommand \@sanitize@url [0]{\catcode `\\12\catcode `\$12\catcode
  `\&12\catcode `\#12\catcode `\^12\catcode `\_12\catcode `\%12\relax}%
\providecommand \@@startlink[1]{}%
\providecommand \@@endlink[0]{}%
\providecommand \url  [0]{\begingroup\@sanitize@url \@url }%
\providecommand \@url [1]{\endgroup\@href {#1}{\urlprefix }}%
\providecommand \urlprefix  [0]{URL }%
\providecommand \Eprint [0]{\href }%
\providecommand \doibase [0]{http://dx.doi.org/}%
\providecommand \selectlanguage [0]{\@gobble}%
\providecommand \bibinfo  [0]{\@secondoftwo}%
\providecommand \bibfield  [0]{\@secondoftwo}%
\providecommand \translation [1]{[#1]}%
\providecommand \BibitemOpen [0]{}%
\providecommand \bibitemStop [0]{}%
\providecommand \bibitemNoStop [0]{.\EOS\space}%
\providecommand \EOS [0]{\spacefactor3000\relax}%
\providecommand \BibitemShut  [1]{\csname bibitem#1\endcsname}%
\let\auto@bib@innerbib\@empty
\bibitem [{\citenamefont {Sigmund}(1969)}]{sigmund69}%
  \BibitemOpen
  \bibfield  {author} {\bibinfo {author} {\bibfnamefont {P.}~\bibnamefont
  {Sigmund}},\ }\href@noop {} {\bibfield  {journal} {\bibinfo  {journal} {Phys.
  Rev.}\ }\textbf {\bibinfo {volume} {184}},\ \bibinfo {pages} {383} (\bibinfo
  {year} {1969})}\BibitemShut {NoStop}%
\bibitem [{\citenamefont {Sigmund}(1973)}]{sigmund73}%
  \BibitemOpen
  \bibfield  {author} {\bibinfo {author} {\bibfnamefont {P.}~\bibnamefont
  {Sigmund}},\ }\href@noop {} {\bibfield  {journal} {\bibinfo  {journal} {J.
  Mater. Sci.}\ }\textbf {\bibinfo {volume} {8}},\ \bibinfo {pages} {1545}
  (\bibinfo {year} {1973})}\BibitemShut {NoStop}%
\bibitem [{\citenamefont {Bradley}\ and\ \citenamefont
  {Harper}(1988)}]{bradley88}%
  \BibitemOpen
  \bibfield  {author} {\bibinfo {author} {\bibfnamefont {R.~M.}\ \bibnamefont
  {Bradley}}\ and\ \bibinfo {author} {\bibfnamefont {J.~M.~E.}\ \bibnamefont
  {Harper}},\ }\href@noop {} {\bibfield  {journal} {\bibinfo  {journal} {J.
  Vac. Sci. Technol. A}\ }\textbf {\bibinfo {volume} {6}},\ \bibinfo {pages}
  {2390} (\bibinfo {year} {1988})}\BibitemShut {NoStop}%
\bibitem [{\citenamefont {Chan}\ and\ \citenamefont {Chason}(2007)}]{chason}%
  \BibitemOpen
  \bibfield  {author} {\bibinfo {author} {\bibfnamefont {W.~L.}\ \bibnamefont
  {Chan}}\ and\ \bibinfo {author} {\bibnamefont {Chason}},\ }\href@noop {}
  {\bibfield  {journal} {\bibinfo  {journal} {J. Appl. Phys.}\ }\textbf
  {\bibinfo {volume} {101}},\ \bibinfo {pages} {121301} (\bibinfo {year}
  {2007})}\BibitemShut {NoStop}%
\bibitem [{\citenamefont {Vasile}\ \emph {et~al.}(1997)\citenamefont {Vasile},
  \citenamefont {Niu}, \citenamefont {Nassar}, \citenamefont {Zhang},\ and\
  \citenamefont {Liu}}]{vasile97}%
  \BibitemOpen
  \bibfield  {author} {\bibinfo {author} {\bibfnamefont {M.~J.}\ \bibnamefont
  {Vasile}}, \bibinfo {author} {\bibfnamefont {Z.}~\bibnamefont {Niu}},
  \bibinfo {author} {\bibfnamefont {R.}~\bibnamefont {Nassar}}, \bibinfo
  {author} {\bibfnamefont {W.}~\bibnamefont {Zhang}}, \ and\ \bibinfo {author}
  {\bibfnamefont {S.}~\bibnamefont {Liu}},\ }\href@noop {} {\bibfield
  {journal} {\bibinfo  {journal} {J. Vac. Sci. Technol. B}\ }\textbf {\bibinfo
  {volume} {15}},\ \bibinfo {pages} {2350} (\bibinfo {year}
  {1997})}\BibitemShut {NoStop}%
\bibitem [{\citenamefont {Adams}\ \emph {et~al.}(2003)\citenamefont {Adams},
  \citenamefont {Vasile}, \citenamefont {Mayer},\ and\ \citenamefont
  {Hodges}}]{adams03}%
  \BibitemOpen
  \bibfield  {author} {\bibinfo {author} {\bibfnamefont {D.}~\bibnamefont
  {Adams}}, \bibinfo {author} {\bibfnamefont {M.}~\bibnamefont {Vasile}},
  \bibinfo {author} {\bibfnamefont {T.}~\bibnamefont {Mayer}}, \ and\ \bibinfo
  {author} {\bibfnamefont {V.}~\bibnamefont {Hodges}},\ }\href@noop {}
  {\bibfield  {journal} {\bibinfo  {journal} {J. Vac. Sci. Technol. B}\
  }\textbf {\bibinfo {volume} {21}},\ \bibinfo {pages} {2334} (\bibinfo {year}
  {2003})}\BibitemShut {NoStop}%
\bibitem [{\citenamefont {Li}\ \emph {et~al.}(2001)\citenamefont {Li},
  \citenamefont {Stein}, \citenamefont {McMullan}, \citenamefont {Branton},
  \citenamefont {Aziz},\ and\ \citenamefont {Golovchenko}}]{li01}%
  \BibitemOpen
  \bibfield  {author} {\bibinfo {author} {\bibfnamefont {J.}~\bibnamefont
  {Li}}, \bibinfo {author} {\bibfnamefont {D.}~\bibnamefont {Stein}}, \bibinfo
  {author} {\bibfnamefont {C.}~\bibnamefont {McMullan}}, \bibinfo {author}
  {\bibfnamefont {D.}~\bibnamefont {Branton}}, \bibinfo {author} {\bibfnamefont
  {M.~J.}\ \bibnamefont {Aziz}}, \ and\ \bibinfo {author} {\bibfnamefont
  {J.~A.}\ \bibnamefont {Golovchenko}},\ }\href@noop {} {\bibfield  {journal}
  {\bibinfo  {journal} {Nature}\ }\textbf {\bibinfo {volume} {412}},\ \bibinfo
  {pages} {166} (\bibinfo {year} {2001})}\BibitemShut {NoStop}%
\bibitem [{\citenamefont {Stein}, \citenamefont {Li},\ and\ \citenamefont
  {Golovchenko}(2002)}]{stein02}%
  \BibitemOpen
  \bibfield  {author} {\bibinfo {author} {\bibfnamefont {D.}~\bibnamefont
  {Stein}}, \bibinfo {author} {\bibfnamefont {J.}~\bibnamefont {Li}}, \ and\
  \bibinfo {author} {\bibfnamefont {J.~A.}\ \bibnamefont {Golovchenko}},\
  }\href@noop {} {\bibfield  {journal} {\bibinfo  {journal} {Phys. Rev. Lett.}\
  }\textbf {\bibinfo {volume} {89}},\ \bibinfo {pages} {276106} (\bibinfo
  {year} {2002})}\BibitemShut {NoStop}%
\bibitem [{\citenamefont {Facsko}\ \emph {et~al.}(1999)\citenamefont {Facsko},
  \citenamefont {Dekorsy}, \citenamefont {Koerdt}, \citenamefont {Trappe},
  \citenamefont {Kurz}, \citenamefont {Vogt},\ and\ \citenamefont
  {Hartnagel}}]{facsko99}%
  \BibitemOpen
  \bibfield  {author} {\bibinfo {author} {\bibfnamefont {S.}~\bibnamefont
  {Facsko}}, \bibinfo {author} {\bibfnamefont {T.}~\bibnamefont {Dekorsy}},
  \bibinfo {author} {\bibfnamefont {C.}~\bibnamefont {Koerdt}}, \bibinfo
  {author} {\bibfnamefont {C.}~\bibnamefont {Trappe}}, \bibinfo {author}
  {\bibfnamefont {H.}~\bibnamefont {Kurz}}, \bibinfo {author} {\bibfnamefont
  {A.}~\bibnamefont {Vogt}}, \ and\ \bibinfo {author} {\bibfnamefont {H.~L.}\
  \bibnamefont {Hartnagel}},\ }\href@noop {} {\bibfield  {journal} {\bibinfo
  {journal} {Science}\ }\textbf {\bibinfo {volume} {285}},\ \bibinfo {pages}
  {1551} (\bibinfo {year} {1999})}\BibitemShut {NoStop}%
\bibitem [{\citenamefont {Frost}, \citenamefont {Schindler},\ and\
  \citenamefont {Bigl}(2000)}]{frost00}%
  \BibitemOpen
  \bibfield  {author} {\bibinfo {author} {\bibfnamefont {F.}~\bibnamefont
  {Frost}}, \bibinfo {author} {\bibfnamefont {A.}~\bibnamefont {Schindler}}, \
  and\ \bibinfo {author} {\bibfnamefont {F.}~\bibnamefont {Bigl}},\ }\href@noop
  {} {\bibfield  {journal} {\bibinfo  {journal} {Phys. Rev. Lett.}\ }\textbf
  {\bibinfo {volume} {85}},\ \bibinfo {pages} {4116} (\bibinfo {year}
  {2000})}\BibitemShut {NoStop}%
\bibitem [{\citenamefont {Cuenat}\ \emph {et~al.}(2005)\citenamefont {Cuenat},
  \citenamefont {George}, \citenamefont {Chang}, \citenamefont {Blakely},\ and\
  \citenamefont {Aziz}}]{cuenat05}%
  \BibitemOpen
  \bibfield  {author} {\bibinfo {author} {\bibfnamefont {A.}~\bibnamefont
  {Cuenat}}, \bibinfo {author} {\bibfnamefont {H.~B.}\ \bibnamefont {George}},
  \bibinfo {author} {\bibfnamefont {K.-C.}\ \bibnamefont {Chang}}, \bibinfo
  {author} {\bibfnamefont {J.}~\bibnamefont {Blakely}}, \ and\ \bibinfo
  {author} {\bibfnamefont {M.~J.}\ \bibnamefont {Aziz}},\ }\href@noop {}
  {\bibfield  {journal} {\bibinfo  {journal} {Adv. Mater.}\ }\textbf {\bibinfo
  {volume} {17}},\ \bibinfo {pages} {2845} (\bibinfo {year}
  {2005})}\BibitemShut {NoStop}%
\bibitem [{\citenamefont {Whitesides}\ and\ \citenamefont
  {Grzybowski}(2002)}]{white}%
  \BibitemOpen
  \bibfield  {author} {\bibinfo {author} {\bibfnamefont {G.}~\bibnamefont
  {Whitesides}}\ and\ \bibinfo {author} {\bibfnamefont {B.}~\bibnamefont
  {Grzybowski}},\ }\href@noop {} {\bibfield  {journal} {\bibinfo  {journal}
  {Science}\ }\textbf {\bibinfo {volume} {295}},\ \bibinfo {pages} {2418}
  (\bibinfo {year} {2002})}\BibitemShut {NoStop}%
\bibitem [{\citenamefont {Holmes-Cerfon}, \citenamefont {Aziz},\ and\
  \citenamefont {Brenner}(2012)}]{MHC2012}%
  \BibitemOpen
  \bibfield  {author} {\bibinfo {author} {\bibfnamefont {M.}~\bibnamefont
  {Holmes-Cerfon}}, \bibinfo {author} {\bibfnamefont {M.}~\bibnamefont {Aziz}},
  \ and\ \bibinfo {author} {\bibfnamefont {M.~P.}\ \bibnamefont {Brenner}},\
  }\href@noop {} {\bibfield  {journal} {\bibinfo  {journal} {Phy. Rev. B}\
  }\textbf {\bibinfo {volume} {85}} (\bibinfo {year} {2012})}\BibitemShut
  {NoStop}%
\bibitem [{\citenamefont {Chen}\ \emph {et~al.}(2005)\citenamefont {Chen},
  \citenamefont {Urquidez}, \citenamefont {Ichim}, \citenamefont {Rodriguez},
  \citenamefont {Brenner},\ and\ \citenamefont {Aziz}}]{chen2005}%
  \BibitemOpen
  \bibfield  {author} {\bibinfo {author} {\bibfnamefont {H.}~\bibnamefont
  {Chen}}, \bibinfo {author} {\bibfnamefont {O.}~\bibnamefont {Urquidez}},
  \bibinfo {author} {\bibfnamefont {S.}~\bibnamefont {Ichim}}, \bibinfo
  {author} {\bibfnamefont {L.}~\bibnamefont {Rodriguez}}, \bibinfo {author}
  {\bibfnamefont {M.}~\bibnamefont {Brenner}}, \ and\ \bibinfo {author}
  {\bibfnamefont {M.}~\bibnamefont {Aziz}},\ }\href {\doibase DOI
  10.1126/science.1117219} {\bibfield  {journal} {\bibinfo  {journal}
  {Science}\ }\textbf {\bibinfo {volume} {310}},\ \bibinfo {pages} {294}
  (\bibinfo {year} {2005})}\BibitemShut {NoStop}%
\bibitem [{\citenamefont {Sigmund}(2006)}]{aziz2006}%
  \BibitemOpen
  \bibinfo {editor} {\bibfnamefont {P.}~\bibnamefont {Sigmund}},\ ed.,\
  \href@noop {} {\emph {\bibinfo {title} {Matematisk-Fysiske Meddelelser / udg.
  af Det Kongelige Danske Videnskabernes Selskab}}},\ \bibinfo {series} {Ion'06
  Proceedings}, Vol.~\bibinfo {volume} {52}\ (\bibinfo {year}
  {2006})\BibitemShut {NoStop}%
\bibitem [{\citenamefont {Mullins}(1959)}]{mullins1959}%
  \BibitemOpen
  \bibfield  {author} {\bibinfo {author} {\bibfnamefont {W.}~\bibnamefont
  {Mullins}},\ }\href@noop {} {\bibfield  {journal} {\bibinfo  {journal} {J.
  Appl. Phys.}\ }\textbf {\bibinfo {volume} {30}} (\bibinfo {year}
  {1959})}\BibitemShut {NoStop}%
\bibitem [{\citenamefont {Herring}(1950)}]{herring1950}%
  \BibitemOpen
  \bibfield  {author} {\bibinfo {author} {\bibfnamefont {C.}~\bibnamefont
  {Herring}},\ }\href@noop {} {\bibfield  {journal} {\bibinfo  {journal} {J.
  Appl. Phys.}\ }\textbf {\bibinfo {volume} {21}} (\bibinfo {year}
  {1950})}\BibitemShut {NoStop}%
\bibitem [{\citenamefont {Umbach}, \citenamefont {Headrick},\ and\
  \citenamefont {Chang}(2001)}]{umbach2001}%
  \BibitemOpen
  \bibfield  {author} {\bibinfo {author} {\bibfnamefont {C.}~\bibnamefont
  {Umbach}}, \bibinfo {author} {\bibfnamefont {R.}~\bibnamefont {Headrick}}, \
  and\ \bibinfo {author} {\bibfnamefont {K.}~\bibnamefont {Chang}},\
  }\href@noop {} {\bibfield  {journal} {\bibinfo  {journal} {Phys. Rev. Lett.}\
  }\textbf {\bibinfo {volume} {87}} (\bibinfo {year} {2001})}\BibitemShut
  {NoStop}%
\bibitem [{\citenamefont {Davidovitch}, \citenamefont {Aziz},\ and\
  \citenamefont {Brenner}(2009)}]{davidovitch2009}%
  \BibitemOpen
  \bibfield  {author} {\bibinfo {author} {\bibfnamefont {B.}~\bibnamefont
  {Davidovitch}}, \bibinfo {author} {\bibfnamefont {M.~J.}\ \bibnamefont
  {Aziz}}, \ and\ \bibinfo {author} {\bibfnamefont {M.~P.}\ \bibnamefont
  {Brenner}},\ }\href@noop {} {\bibfield  {journal} {\bibinfo  {journal} {J.
  Phys.: Condens. Matter}\ }\textbf {\bibinfo {volume} {21}} (\bibinfo {year}
  {2009})}\BibitemShut {NoStop}%
\bibitem [{\citenamefont {Strikwerda}(2004)}]{strikwerda}%
  \BibitemOpen
  \bibfield  {author} {\bibinfo {author} {\bibfnamefont {J.}~\bibnamefont
  {Strikwerda}},\ }\href@noop {} {\emph {\bibinfo {title} {Finite difference
  schemes and partial differential equations}}},\ \bibinfo {edition} {2nd}\
  ed.\ (\bibinfo  {publisher} {SIAM},\ \bibinfo {year} {2004})\BibitemShut
  {NoStop}%
\bibitem [{\citenamefont {Bertozzi}\ \emph {et~al.}(1993)\citenamefont
  {Bertozzi}, \citenamefont {Brenner}, \citenamefont {Dupont},\ and\
  \citenamefont {Kadanoff}}]{bertozzi93}%
  \BibitemOpen
  \bibfield  {author} {\bibinfo {author} {\bibfnamefont {A.}~\bibnamefont
  {Bertozzi}}, \bibinfo {author} {\bibfnamefont {M.}~\bibnamefont {Brenner}},
  \bibinfo {author} {\bibfnamefont {T.}~\bibnamefont {Dupont}}, \ and\ \bibinfo
  {author} {\bibfnamefont {L.}~\bibnamefont {Kadanoff}},\ }in\ \href@noop {}
  {\emph {\bibinfo {booktitle} {Trends and Perspectives in Applied
  Mathematics}}},\ Vol.\ \bibinfo {volume} {100},\ \bibinfo {editor} {edited
  by\ \bibinfo {editor} {\bibfnamefont {L.}~\bibnamefont {Sirovich}}}\
  (\bibinfo  {publisher} {Springer-Verlag},\ \bibinfo {year} {1993})\ pp.\
  \bibinfo {pages} {155--208}\BibitemShut {NoStop}%
\bibitem [{\citenamefont {Vollmary-Lee}\ and\ \citenamefont
  {Rutenberg}(2003)}]{vollmary2003}%
  \BibitemOpen
  \bibfield  {author} {\bibinfo {author} {\bibfnamefont {B.}~\bibnamefont
  {Vollmary-Lee}}\ and\ \bibinfo {author} {\bibfnamefont {A.}~\bibnamefont
  {Rutenberg}},\ }\href@noop {} {\bibfield  {journal} {\bibinfo  {journal}
  {Phys. Rev. E}\ }\textbf {\bibinfo {volume} {68}} (\bibinfo {year}
  {2003})}\BibitemShut {NoStop}%
\bibitem [{\citenamefont {Gao}\ and\ \citenamefont {Bertozzi}(2011)}]{gao2011}%
  \BibitemOpen
  \bibfield  {author} {\bibinfo {author} {\bibfnamefont {W.}~\bibnamefont
  {Gao}}\ and\ \bibinfo {author} {\bibfnamefont {A.}~\bibnamefont {Bertozzi}},\
  }\href@noop {} {\bibfield  {journal} {\bibinfo  {journal} {{SIAM J. Imag.
  Sci.}}\ }\textbf {\bibinfo {volume} {4}},\ \bibinfo {pages} {597} (\bibinfo
  {year} {2011})}\BibitemShut {NoStop}%
\bibitem [{\citenamefont {Schonlieb}\ and\ \citenamefont
  {Bertozzi}(2011)}]{carola2011}%
  \BibitemOpen
  \bibfield  {author} {\bibinfo {author} {\bibfnamefont {C.-B.}\ \bibnamefont
  {Schonlieb}}\ and\ \bibinfo {author} {\bibfnamefont {A.}~\bibnamefont
  {Bertozzi}},\ }\href@noop {} {\bibfield  {journal} {\bibinfo  {journal}
  {Comm. Math. Sci.}\ }\textbf {\bibinfo {volume} {9}},\ \bibinfo {pages} {413}
  (\bibinfo {year} {2011})}\BibitemShut {NoStop}%
\bibitem [{\citenamefont {Bertozzi}, \citenamefont {Ju},\ and\ \citenamefont
  {Lu}(2011)}]{bertozzi2011}%
  \BibitemOpen
  \bibfield  {author} {\bibinfo {author} {\bibfnamefont {A.}~\bibnamefont
  {Bertozzi}}, \bibinfo {author} {\bibfnamefont {N.}~\bibnamefont {Ju}}, \ and\
  \bibinfo {author} {\bibfnamefont {H.}~\bibnamefont {Lu}},\ }\href@noop {}
  {\bibfield  {journal} {\bibinfo  {journal} {Discrete Contin. Dyn. S.}\
  }\textbf {\bibinfo {volume} {29}},\ \bibinfo {pages} {1367} (\bibinfo {year}
  {2011})}\BibitemShut {NoStop}%
\bibitem [{\citenamefont {Yamamura}, \citenamefont {Itikawa},\ and\
  \citenamefont {Itoh}(1983)}]{yamamura1983}%
  \BibitemOpen
  \bibfield  {author} {\bibinfo {author} {\bibfnamefont {Y.}~\bibnamefont
  {Yamamura}}, \bibinfo {author} {\bibfnamefont {Y.}~\bibnamefont {Itikawa}}, \
  and\ \bibinfo {author} {\bibfnamefont {N.}~\bibnamefont {Itoh}},\ }\href@noop
  {} {\enquote {\bibinfo {title} {Angular dependence of sputtering yields of
  monatomic solids},}\ }\bibinfo {howpublished} {Report No. IPPJ-AM-26}
  (\bibinfo {year} {1983})\BibitemShut {NoStop}%
\end{thebibliography}%

\end{document}